\documentclass[]{jaa}

\usepackage{graphicx}
\usepackage[round]{natbib}
\usepackage{multicol}

\def \aj {AJ}
\def \mnras {MNRAS}
\def \apj {ApJ}

\def \apss {Ap\&SS}
\def \apjl {ApJL}
\def \aap {A\&A}
\def \nat {Nature}
\def \araa {ARAA}

\def \pasj {PASJ}

\def \prc {PhRvC}

\def \pre {PhRvE}

\begin{document}

\title{Cooling of accretion-heated neutron stars}


\author{Rudy Wijnands\textsuperscript{1,*}, Nathalie Degenaar\textsuperscript{1} \and Dany Page\textsuperscript{2}}
\affilOne{\textsuperscript{1}Anton Pannekoek Institute for Astronomy, University of Amsterdam, Postbus 94249, 1090 GE Amsterdam, The Netherlands\\}{
\affilTwo{\textsuperscript{2}Instituto de Astronom\'\i a, Universidad Nacional Aut\'onoma de M\'exico, M\'exico D.F., 04510, M\'exico}


\twocolumn[{

\maketitle

\corres{r.a.d.wijnands@uva.nl}


\begin{abstract}
We present a brief, observational review about the study of the cooling behaviour of accretion-heated neutron stars and the inferences about the neutron-star crust and core that have been obtained from these studies. Accretion of matter during outbursts can heat the crust out of thermal equilibrium with the core and after the accretion episodes are over, the crust will cool down until crust-core equilibrium is restored. We discuss the observed properties of the crust cooling sources and what has been learned about the physics of neutron-star crusts. We also briefly discuss those systems that have been observed long after their outbursts were over, i.e, during times when the crust and core are expected to be in thermal equilibrium. The surface temperature is then a direct probe for the core temperature. By comparing the expected temperatures based on estimates of the accretion history of the targets with the observed ones, the physics of neutron-star cores can be investigated. Finally, we discuss similar studies performed for strongly magnetized neutron stars in which the magnetic field might play an important role in the heating and cooling of the neutron stars.
\end{abstract}


}]


\doinum{10.1007/s12036-017-9466-5}
\artcitid{49}
\volnum{38}
\year{September 2017}
\pgrange{1–16}
\setcounter{page}{1}
\lp{16}

\section{Introduction}

Neutron stars (NSs) in X-ray binaries accrete matter from their companion stars. Several types of X-ray binaries can be identified. In low-mass X-ray binaries (LMXBs), the companion has a mass lower than that of the NS so that stable Roche-lobe overflow can occur. In high-mass X-ray binaries (HMXBs), matter is transferred either through the strong stellar wind of the massive donor or through the decretion disk of a Be-type companion star.  In LMXBs the NSs typically have low magnetic field strengths ($B \sim 10^{8-9}$ G), while in HMXBs the NSs are more strongly magnetized ($B \sim 10^{12-13}$ G). 

Most X-ray binaries do not accrete persistently at high mass-accretion rates. Many systems are X-ray transients and they only accrete episodically during sporadic outbursts. The quiescent periods can last years to decades during which the systems do not accrete or accrete only at very low rates resulting in very low quiescent luminosities ($L_{\rm q} \sim 10^{31-34}$ erg s$^{-1}$). During the outbursts the sources are much more luminous (with X-ray luminosities $> 10^{36-39}$ ergs s$^{-1}$)  due to the very high accretion rates. Such outbursts typically last for weeks to months (the ``ordinary'' transients), but some systems stay active for years to even decades (the ``quasi-persistent'' transients).

In quiescence, X-ray transients are very faint and can only be studied in detail with sensitive X-ray satellites. In this short, observational review we will focus on the quiescent behaviour of LMXBs and what can be learned from those systems about NS physics, although in Section \ref{section:magnetic} we will briefly discuss the HMXBs as well\footnote{We will also not discuss the candidate quiescent NS LMXBs in globular clusters \citep[see, e.g., the following recent publications:][]{2012MNRAS.423.1556S,2013ApJ...764..145C,2014MNRAS.444..443H,2016ApJ...831..184B}. The main reason is that no X-ray outbursts have been observed for these sources.}. Some NS LMXBs could already be detected and studied  using the X-ray telescopes on board {\it Einstein}, {\it EXOSAT}, {\it ROSAT}, {\it ASCA}, and {\it BeppoSAX} \citep[see Table \ref{table:qsources} and the review by][]{1998A&ARv...8..279C}. However, those early studies focussed on the closest and brightest systems in quiescence (i.e., Aql X-1, Cen X-4, 4U 1608-52, EXO 0748-676, 4U 2129+47, and the Rapid Burster\footnote{It still remains to be determined if the quiescent source studied by \citet[][]{1996PASJ...48L..27A,1996PASJ...48..257A} is indeed the Rapid Burster and not an unrelated source in the same globular cluster (Liller 1). In addition, some of these other cluster sources might have contributed to the inferred flux of the Rapid Burster even if most of it originated from this source.}). The study of quiescent transients fully matured when {\it Chandra} and {\it XMM-Newton} were launched in 1999, followed in 2004 by the launch of {\it Swift}. The studies of quiescent systems performed with these satellites first focussed on the already studied systems, but later also the less well known systems were included (as well as newly discovered systems). We will not discuss all sources individually, but we will only focus on the general properties observed from these sources.  In Tables \ref{table:qsources}-\ref{table:BeX} we list the sources that have been observed in quiescence as well as the articles reporting those observations. These tables can be used as a starting point if more detailed information about the individual sources is required.

In general, quiescent NS LMXBs have  $L_{\rm q} <10^{33-34}$ erg s$^{-1}$ (0.5-10 keV) and their quiescent X-ray spectra can be well described using a combination of a soft component (dominating below 1 keV)  and a hard component (dominating above 3 keV). The soft component is typically modelled using a NS atmosphere model and the hard component using a simple power-law model. For some systems their quiescent spectra are fully dominated by the soft component \citep[e.g.,][]{2001ApJ...559.1054R,2004ApJ...610..933T}, while for others their spectra are totally dominated by the hard component \citep[e.g.,][]{2002ApJ...575L..15C,2005ApJ...618..883W,2012ApJ...756..148D}.

The soft component likely originates from the surface of the NS and could be due to cooling emission from a hot NS that has been heated due to the accretion of matter (Section \ref{section:heating}). Alternatively, it could be due to residual accretion of matter onto the surface of the NS. The power-law component is not well understood \citep[see, e.g.,][]{1998A&ARv...8..279C}. It could arise from residual accretion as well \citep[see, e.g., the discussions in][]{2014ApJ...797...92C,2015MNRAS.449.2803D,2015MNRAS.454.1371W} or it could be related to processes involving the magnetosphere of the NS \citep[for a recent discussion see][]{2012ApJ...756..148D}. Since the topic of this review is the study of the cooling of accretion-heated NSs, we will not further discuss other mechanisms. It is usually assumed that when the quiescent spectra are dominated by the soft component, we indeed observe cooling from a hot NS. However, when the power-law component contributes significantly (i.e., it contributes more than several tens of percent to the quiescent luminosity that is observed) then the cooling emission from the NS will be very likely contaminated by this extra component. In those cases,  the inferred surface temperatures and quiescent luminosities should be regarded as upper limits.

\begin{figure*}[!th]
\includegraphics[width=1.45\columnwidth,angle=90]{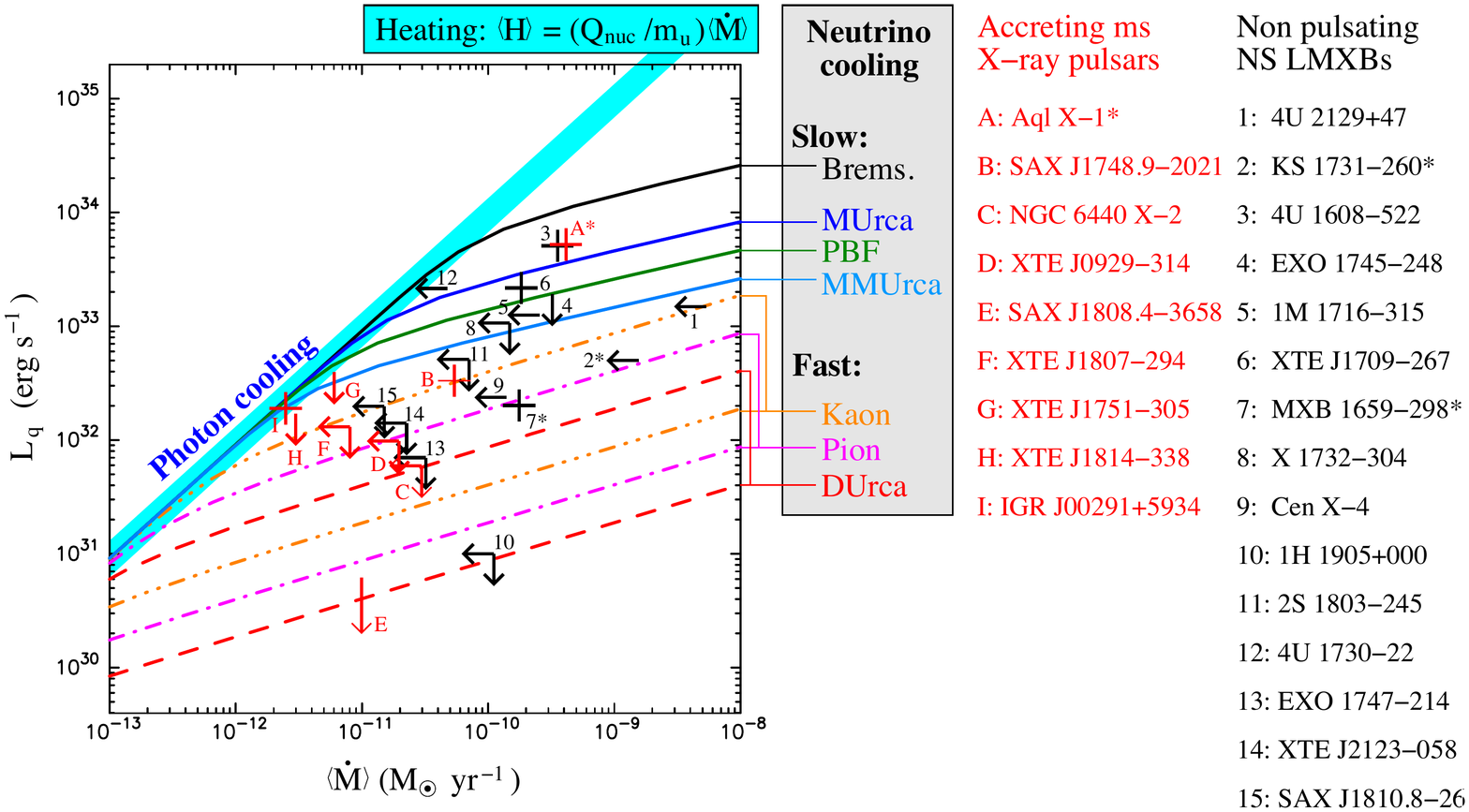}
\vspace{-1.2cm}
\caption{The $L_{\rm q}$ versus $\langle \dot{M} \rangle$ for a sample of transient NS LMXBs as well as several cooling scenarios \citep[adapted from][]{2013MNRAS.432.2366W}. The red points are AMXPs (Table \ref{table:qAMXPs}) and black points are non-pulsating NS LMXBs (Table \ref{table:qsources}). The data points are from \citet[][]{2010ApJ...714..894H}. The systems that are indicated with an * are systems for which crust cooling has been observed (Section \ref{section:sources}) and the displayed $L_{\rm q}$ should be regarded as upper limits. The blue ``Heating'' band shows the predicted values for the average heating rate $\langle H \rangle$ (with $Q_{\rm nuc}$ the amount of heat deposited in the crust per accreted nucleon and $m_{\rm u}$ the atomic mass unit).  The text ``Photon cooling'' indicates where the cooling due to thermal emission from the surface of the NS equals the heat that has been generated. The bands indicated for the fast cooling processes demonstrate the large uncertainties in our understanding of these processes. }\label{fig:lxversusmdot}
\end{figure*}

\begin{table*}[!htb]
\fontsize{8}{10}\selectfont
\caption{Reports on observations of quiescent non-pulsating neutron-star LMXBs}\label{table:qsources}
\begin{tabular}{lccl}
\topline
Sources                & Instrument$^a$ & Remark      & References\\\midline
EXO 0748-676$^b$       & E, X           & Eclipser    & \citet[][]{1986ApJ...308..199P,1999AJ....118.1390G}\\
                       &                &             & \citet[][]{2011MNRAS.414.1077Z,2017arXiv170604784C}\\
SAX J1324.5-6313       & C              &             & \citet[][]{2002AA...392..931C}\\
MAXI J1421-613         & C              &             & \citet[][]{2014ATel.5894....1C}\\
Cen X-4 (4U 1456-32)   & A, B, C, EX,   &             & \citet[][]{1987AA...182...47V,1994ApJ...435..407G,1996PASJ...48..257A,1998PASJ...50..611A}\\
                       & N, R, S, X     &             & \citet[][]{1997AA...324..941C,2000AA...358..583C,1999ApJ...514..945R,2001ApJ...551..921R}\\
                       &                &            & \citet[][]{2010ApJ...720.1325C,2013MNRAS.433.1362C,2013MNRAS.436.2465B}\\
                       &                &             & \citet[][]{2014ApJ...797...92C,2015MNRAS.449.2803D}\\
4U 1608-522            & A              &             & \citet[][]{1996PASJ...48..257A,1998PASJ...50..611A,1999ApJ...514..945R}\\
XTE J1701-407          & S              &             & \citet[][]{2011ATel.3604....1D}\\
XTE J1709-267          & C              &             & \citet[][]{2003MNRAS.341..823J,2004MNRAS.354..666J}\\
2S 1711-339            & C              &             & \citet[][]{2004ATel..233....1T,2004ATel..238....1T}\\
1M 1716-315            & C              &             & \citet[][]{2007MNRAS.377.1295J}\\
IGR J17191-2821        & C              &             & \citet[][]{2007ATel.1096....1C}\\
XTE J1723-376          & C              &             & \citet[][]{2008MNRAS.391L.117B}\\
4U 1730-22             & C              &             & \citet[][]{2007ApJ...663..461T}\\
Rapid Burster (4U 1730-335)          & A              & In Liller 1 & \citet[][]{1996PASJ...48L..27A,1998PASJ...50..611A}\\
KS 1731-260            & B              &             & \citet[][]{2002ApJ...574..930B}\\
X 1732-304             & C              & In Terzan 1 & \citet[][]{2002ApJ...572.1002W,2006MNRAS.369..407C}\\
Swift J1734.5-3027     & S              &             & \citet[][]{2015AA...579A..56B}\\
IGR J17445-2747        & C, X           &             & \citet[][]{2008ApJ...685.1143T,2010MNRAS.408..975M}\\ 
EXO 1745-248           & C              & In Terzan 5 & \citet[][]{2005ApJ...618..883W,2006ApJ...651.1098H}\\
                       &                &             & \citet[][]{2012MNRAS.422..581D}\\
EXO 1747-214           & C              &             & \citet[][]{2005ApJ...635.1233T}\\
GRS 1747-312           & X              & In Terzan 6 & \citet[][]{vats}\\
                       &                & Eclipser    & \\
XMMU J174716.1-281048  & C, X           &             & \citet[][]{2007AA...468L..17D}\\
Swift J174805.3-244637$^b$ & C          & In Terzan 5 & \citet[][]{2014ApJ...780..127B}\\
SAX J1750.8-2900       & C, S, X        &             & \citet[][]{2012ApJ...749..111L,2013MNRAS.434.1599W}\\
                       &                &             & \citet[][]{parikhwijnands}\\
SAX J1752.3-3128       & C              &             & \citet[][]{2002AA...392..931C}\\
SAX J1753.5-2349       & C              &             & \citet[][]{2002AA...392..931C}\\
AX J1754.2-2754        & C              &             & \citet[][]{2008ATel.1575....1B}\\
2S 1803-245            & X              &             & \citet[][]{2007MNRAS.380.1637C}\\
SAX J1806.5-2215       & C              &             & \citet[][]{2002AA...392..931C}\\
SAX J1810.8-2609       & C              &             & \citet[][]{2004MNRAS.349...94J}\\
SAX J1818.7+1424       & C              &             & \citet[][]{2002AA...392..931C}\\
SAX J1828.5-1037       & S              &             & \citet[][]{2009ApJ...699.1144C}\\
Swift J185003.2-005627 & X              &             & \citet[][]{2012ApJ...759....8D}\\
1H 1905+000            & C              &             & \citet[][]{2006MNRAS.368.1803J,2007ApJ...665L.147J}\\
Swift J1922.7-1716     & S              &             & \citet[][]{2012ApJ...759....8D}\\
XTE J2123-058          & C              &             & \citet[][]{2004ApJ...610..933T}\\
M15 X-3$^c$            & C, R           & in M15      & \citet[][]{2009ApJ...692..584H}\\
4U 2129+47             & C, R, X        & Eclipser    & \citet[][]{1994ApJ...435..407G,1999AJ....118.1390G,2000ApJ...529..985R}\\
                       &                &             & \citet[][]{2002ApJ...573..778N,2007AA...476..301B,2009ApJ...706.1069L}\\
SAX J2224.9+5421       & S, X           &             & \citet[][]{2009ApJ...699.1144C,2014ApJ...787...67D}\\
\hline
\multicolumn{2}{l}{\underline {Sources near Sgr A$^*$}$^d$} & &\\
GRS 1741.9-2853        & C, S, X        &             & \citet[][]{2003ApJ...598..474M}\\
KS 1741-293            & C, S, X        &             & \citet[][]{2013IAUS..290..113D}\\
XMM J174457-2850.3     & C, S, X        &             & \citet[][]{2005MNRAS.357.1211S,2014ApJ...792..109D}\\
AX J1745.6-2901        & C, S, X        & Eclipser    & \citet[][]{2015MNRAS.446.1536P}\\ 
SAX J1747.0-2853       & C              &             & \citet[][]{2012AA...545A..49D}\\ 
\hline
\end{tabular}
\tablenotes{$^a$ A: {\it ASCA}; B: {\it BeppoSAX}; C: {\it Chandra}; E: {\it Einstein}; EX: {\it EXOSAT}; N: {\it NuSTAR}; R: {\it ROSAT}; S: {\it Swift}; X: {\it XMM-Newton} }
\vspace{-0.4cm}
\tablenotes{$^b$ Only non crust cooling references are given. For the crust cooling references see Table \ref{table:cooling}.}
\vspace{-0.4cm}
\tablenotes{$^c$ Although M15 X-3 likely harbours a neutron star, this needs to be confirmed.}
\vspace{-0.4cm}
\tablenotes{$^d$ This field has been observed by many X-ray satellites. Only source specific papers are listed here. Additional references to survey papers: \citet[][]{2006AA...449.1117W,2009AA...495..547D,2010AA...524A..69D,2012AA...545A..49D,2015MNRAS.453..172P}.}
\end{table*}

\section{Heating and cooling of the NS} \label{section:heating}

When matter falls onto the NS surface during accretion outbursts, heat will be generated. Firstly, large amounts of gravitational energy will be released when the matter falls onto the surface causing the surface temperatures to rise to $\sim10^7$ K. If the NS core temperature is lower than this value, some of this heat might flow inwards. However, compression of material at the bottom of the ocean results in electron capture reactions that rapidly heat up these layers to temperatures well above the surface temperature. Therefore, the inflow of the heat generated by gravitational energy release will be halted \citep[e.g.,][]{1984ApJ...278..813F,1990ApJ...362..572M}. If the temperature gradient from the ocean towards the deeper layers in the crust is still negative, a fraction of the heat generated by these thermonuclear reactions can still flow inward, heating up the NS core.

\begin{table*}[htb]
\fontsize{8}{10}\selectfont
\caption{Reports on observations of quiescent accreting millisecond X-ray pulsars$^a$}\label{table:qAMXPs}
\begin{tabular}{lccl}
\topline
Sources                & Instrument$^b$ & Remark      & References\\\midline
IGR J00291+5934        & C, X           &             & \citet[][]{2005MNRAS.361..511J,2008ApJ...680..615J,2008ApJ...672.1079T,2008ApJ...689L.129C}\\
                       &                &             & \citet[][]{2009ApJ...691.1035H}\\
Swift J0911.9-6453     & C              & In NGC 2808 & \citet[][]{2016ATel.8971....1H,2008AA...490..641S}\\
XTE J0929-314          & C, X           &             & \citet[][]{2005ApJ...619..492W,2005AA...434L...9C}\\
GRO J1744-28$^c$       & C, X           & 2.1 Hz pulsar & \citet[][]{2002ApJ...568L..93W,2002AA...386..531D,2012AA...545A..49D}\\
IGR J17480-2446$^{c,d}$ & C              & 11 Hz pulsar& \citet[][]{2011MNRAS.412L..68D}\\
                       &                & In Terzan 5 & \\ 
SAX J1748.9-2021       & C              & In NGC 6440 & \citet[][]{2001ApJ...563L..41I,2005ApJ...620..922C,2015MNRAS.449.1238W}\\
                       &                &             & \citet[][]{2015MNRAS.452.3475B}\\
NGC 6440 X-2           & C              & In NGC 6440 & \citet[][]{2010ApJ...714..894H}\\
Swift J1749.4-2807     & S, X           &             & \citet[][]{2009MNRAS.393..126W,2009ApJ...699.1144C,2012ApJ...756..148D}\\
IGR J17498-2921        & C              &             & \citet[][]{2011ATel.3559....1J}\\
XTE J1751-305          & C              &             & \citet[][]{2005ApJ...619..492W}\\
IGR J17511-3057        & S              &             & \citet[][]{2012MNRAS.424...93H}\\
Swift J1756.9-2508     & C              &             & \citet[][]{2007ATel.1133....1P}\\
XTE J1807-294          & X              &             & \citet[][]{2005AA...434L...9C}\\
SAX J1808.4-3658       & A, B, X        &             & \citet[][]{2000ApJ...537L.115S,2000ApJ...543L.145D,2002ApJ...571..429W}\\
                       &                &             & \citet[][]{2002ApJ...575L..15C,2007ApJ...660.1424H,2009ApJ...691.1035H}\\
XTE J1814-338          & X              &             & \citet[][]{2009ApJ...691.1035H}\\
IGR J18245-2452        & C              & In M28      & \citet[][]{2014MNRAS.438..251L}\\
Aql X-1$^d$ (4U 1908+00)           & A, B, C,       &           & \citet[][]{1987AA...182...47V,1994AA...285..903V,1998PASJ...50..611A}\\
                       & EX, R, X       &             & \citet[][]{1998ApJ...499L..65C,1999ApJ...514..945R,2001ApJ...559.1054R,2002ApJ...577..346R}\\
                       &                &             & \citet[][]{2003ApJ...597..474C,2011MNRAS.414.3006C,2017ApJ...845....8L}\\
\hline
\end{tabular}
\tablenotes{$^a$ A similar table was also presented by \citet[][]{2012ApJ...756..148D}.}
\vspace{-0.4cm}
\tablenotes{$^b$ A: {\it ASCA}; B: {\it BeppoSAX}; C: {\it Chandra}; EX: {\it EXOSAT}; R: {\it ROSAT}; S: {\it Swift}; X: {\it XMM-Newton} }
\vspace{-0.4cm}
\tablenotes{$^c$ Per definition, GRO J1744-28 and IGR J17480-2446 are not AMXPs but we list them here for completeness as well.}
\vspace{-0.4cm}
\tablenotes{$^d$ Only non crust cooling references are given. For the crust cooling references see Table \ref{table:cooling}.}
\end{table*}

The NS core is most likely more significantly heated by nuclear reactions occurring deeper in the crust \citep[][]{1998ApJ...504L..95B}. When the matter falls onto the NS, the underlying, previously accreted matter is compressed to deeper layers and thus higher densities. Eventually, the original catalysed crust will be replaced by a crust made out of accreted matter \citep[][the original crust will have fused with the core]{1979PThPh..62..957S,1990A&A...229..117H}. Such an accreted crust will have a completely different composition than the original crust; i.e., it will be richer in low-Z elements. In such a crust, the compression of the matter will induce a variety of nuclear reactions, such as electron captures, neutron emission, and pycnonuclear reactions \citep[e.g.,][]{1979PThPh..62..957S,1990A&A...227..431H,2003A&A...404L..33H,2008A&A...480..459H}. In total, those reactions can generate about $\sim$1-2 MeV per accreted nucleon. Most energy is generated by the pycnonuclear reactions that occur in the layers of the crust that have densities of $10^{12-13}$ g cm$^{-3}$ \citep[therefore this process is often called the ``deep crustal heating'' model;][]{1998ApJ...504L..95B}. These layers are heated up by these reactions and the produced heat slowly flows into the stellar core. 

\citet[][]{1998ApJ...504L..95B} demonstrated that the core is heated on a (thermal) time scale of $\sim10^{4}$ years \citep[see also][]{2001ApJ...548L.175C,2013MNRAS.432.2366W}. In quiescence, this heat would be transported to the surface where it would be released as thermal emission, cooling down the NS. The observed quiescent luminosity will depend on the time-averaged accretion rate ($\langle \dot{M} \rangle$; averaged over the thermal time scale) onto the NS and the neutrino emission processes active in the core. To test the deep crustal heating model, the predictions for for the quiescent luminosity as a function of $\langle \dot{M} \rangle$ for the various neutrino emission mechanisms are compared with the observed luminosities using diagrams similar to Figure  \ref{fig:lxversusmdot} \citep[after][]{2004ARA&A..42..169Y,2007ApJ...660.1424H,2009ApJ...691.1035H,2010ApJ...714..894H,2013MNRAS.432.2366W,2017arXiv170208452H}. In this figure, the non-pulsating NS LMXBs (Table \ref{table:qsources}) are plotted separately from the accreting millisecond X-ray pulsars (AMXPs; Table \ref{table:qAMXPs}) to investigate if the pulsars behave different than the non-pulsating sources with respect to their heating and cooling behaviour. However, so far no clear difference has been found, although the AMXPs tend to be, on average, colder than the non-pulsating sources but this could simply be due to the differences in the $\langle \dot{M} \rangle$ between source classes (and not related to internal NS physics).

The cooling mechanisms displayed in Figure \ref{fig:lxversusmdot} can roughly be divided in slow (standard) neutrino cooling processes and fast (enhanced) mechanisms. These processes differ significantly in their efficiency and temperature dependence \citep[for a detailed discussion see][]{2004ARA&A..42..169Y}.  A full description of these processes in relation to Figure~\ref{fig:lxversusmdot} is given by \citet[][]{2013MNRAS.432.2366W}. In the remainder of this paragraph, we only briefly summarise the most relevant information concerning these mechanisms. The slow emission mechanisms include several bremsstrahlung processes (called ``Brems.''~in Fig.~\ref{fig:lxversusmdot}) and the modified Urca (``MUrca'') process. A more efficient cooling process is the Medium-Modified Urca process \citep[``MMurca'';][]{MMUrca}, which might be active if the pion mode softens (a possible precursor to the pion condensate). The most efficient fast process is the direct Urca (``DUrca'') process involving nucleons. If hyperons are present, their related DUrca processes are also active although with lower efficiencies.  If deconfined quark matter is present in the inner core of the NS, additional DUrca processes are present with efficiencies that can be as large as the nucleon DUrca one. If a meson condensate (either kaon or pion) is present in the core, other fast, emission processes are possible, although they are less efficient than the DUrca processes. If Cooper pairing of the nucleons occur (causing, e.g., neutron superfluidity), the emission processes (both the slow and the fast) are strongly reduced \citep[e.g.,][]{2006NuPhA.777..497P}, although it opens also a new neutrino emission mechanism due to the constant formation and breaking of the pairs (``pair breaking and formation'' or ``PBF'').

From Figure~\ref{fig:lxversusmdot} it can be concluded that some systems are consistent with slow neutrino cooling processes, but that most systems need fast neutrino emission mechanisms to cool the core to the observed temperature.  Such enhanced neutrino emission processes are expected when the core density has reached a certain threshold and thus the cold systems might harbour relatively massive NSs \citep[e.g.,][]{2001ApJ...548L.175C,2004ARA&A..42..169Y,2013MNRAS.432.2366W}. However, significant uncertainties are present in this diagram. The quiescent luminosities have not be determined in a homogenous way and some sources (e.g., Aql X-1, Cen X-4, KS 1731-260, MXB 1659-29) are variable in quiescence and not always the lowest luminosity values have been used. In addition, the $\langle \dot{M} \rangle$ must be averaged over the thermal time scale of the NS cores ($\sim 10^{3-4}$ years) but observationally 
$\langle \dot{M} \rangle$ can only be determined over the last few decades (i.e., the time since the birth of X-ray astronomy). This short term $\langle \dot{M} \rangle$ could be considerably different than the one needed for this figure. Finally, the distance is often not well known but since this affects both the quiescent luminosities as $\langle \dot{M} \rangle$, this will result in a diagonal shift in this diagram \citep[][]{2007ApJ...660.1424H,2009ApJ...691.1035H,2010ApJ...714..894H}.

\begin{figure*}[t]
\includegraphics[width=\columnwidth]{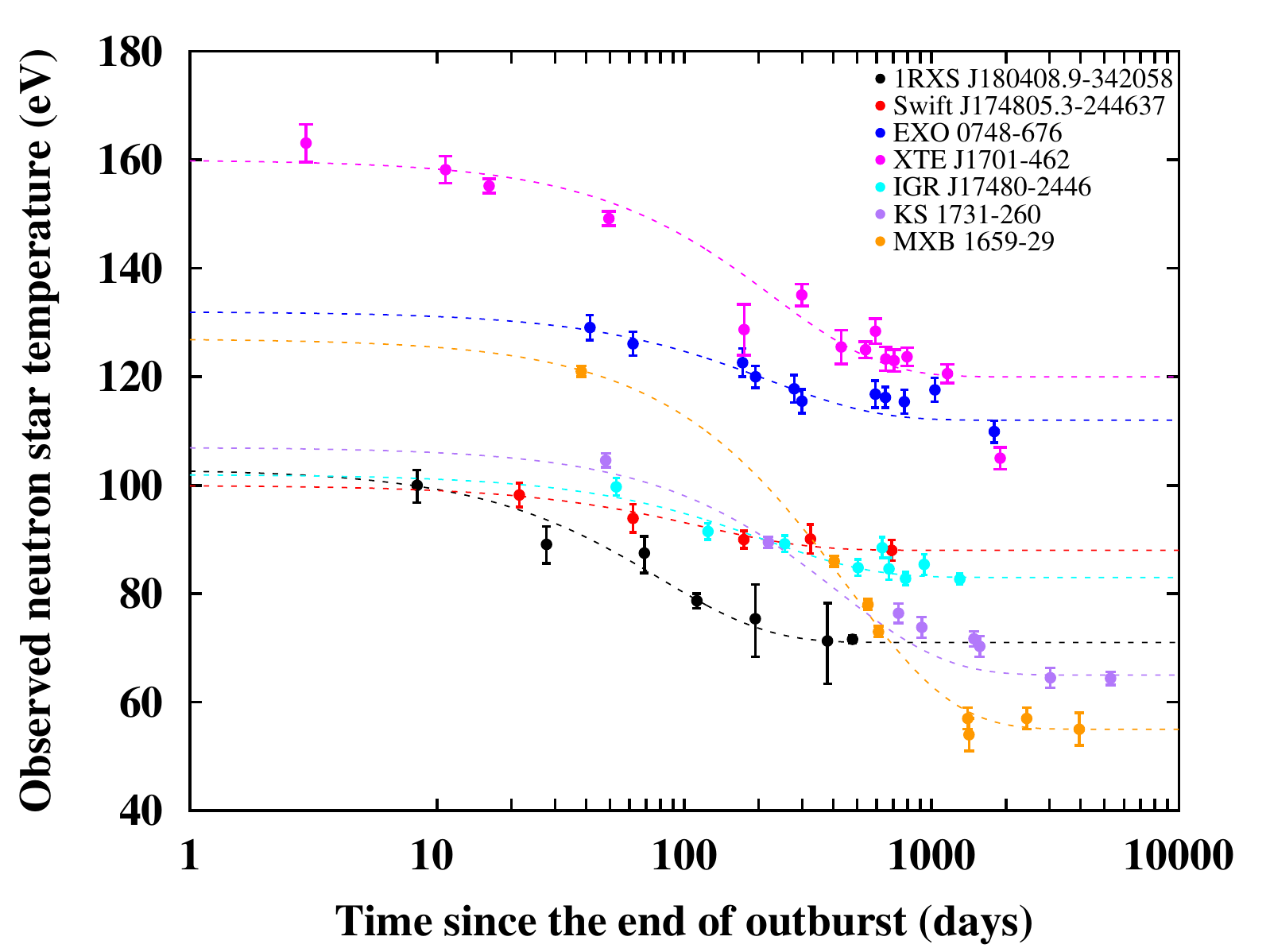}
\includegraphics[width=\columnwidth]{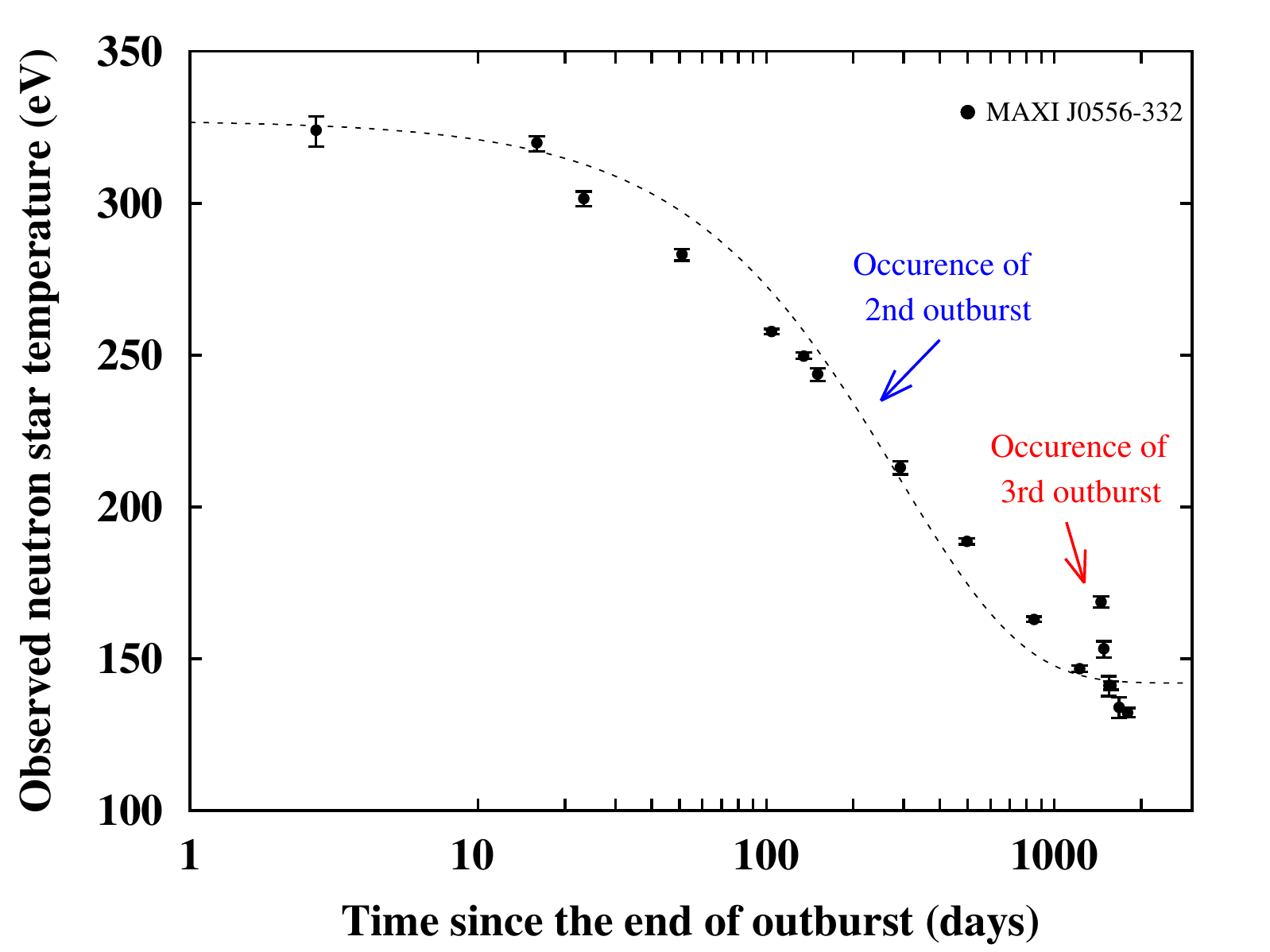}
\caption{{\bf Left:} The crust cooling curves for all monitored systems except MAXI J0556-332 since this source has a very hot surface temperature and including this source in this figure would inhibit to see clearly the curves obtained for the other sources. {\bf Right:} The crust cooling curve of MAXI J0556-332, also showing (after day 1000) the reheating of the crust (followed again by subsequent cooling) that occurred during the 2016 outburst of the source \citep[adapted from][]{parikh0556}. The dotted curves through the data points (in both figures) is a simple exponential decay function that levels off at a certain constant value. This curve is just to guide the eye for the individual sources and it is clear that this simple mathematical model cannot explain the observed cooling curves.  For both XTE J1701-462 and MAXI J0556-332 several brief and weak accretion flares were seen during the cooling phase \citep[][]{2010ApJ...714..270F,2011ApJ...736..162F,2014ApJ...795..131H}. Those flares have been removed from the data.  }\label{fig:coolingcurves}
\end{figure*}

Besides the systematics in the observables, the models also have their own uncertainties. We refer to \citet[][]{2013MNRAS.432.2366W} and \citet[][]{2017arXiv170208452H} for discussions. One main uncertainty is the unknown composition of the envelope, which can be quite different between sources \citep[][]{2002ApJ...574..920B}. From observations, the surface temperature is obtained and from that the core temperature is inferred using an envelope model \citep[e.g.,][]{1997A&A...323..415P}. However, for a particular observed temperature, the inferred core temperature is lower if the envelope contains more light elements than when the envelope contains significant amounts of heavy elements. The  effects of this uncertainty were recently studied by \citet[][]{2017arXiv170208452H}. Despite the  uncertainties in the observations and the models, it is unlikely that all sources would shift to the standard, slow cooling track and the general conclusions inferred from this figure are still valid.

\section{Crust cooling sources}\label{section:sources}

One of the main assumptions behind the studies presented in Section \ref{section:heating} is that the crust and core are in thermal equilibrium during the observations in the quiescent phase. This is a reasonable assumption if a system is observed long after the end of its last outburst. This would allow the heat generated in the crust to be fully transported inwards to the core and outwards to the surface. For ordinary transients that have short outbursts (weeks to months), it  was long assumed that this was indeed a valid assumption because of the relatively small expected increase in the crust temperature during the outburst. Very quickly after the outburst ends the crust and core would be in equilibrium again  \citep[e.g.,][but see Section \ref{section:ordinary}]{1998ApJ...504L..95B,2001MNRAS.325.1157U,2002ApJ...580..413R}. However, the situation is different for the quasi-persistent transients. The very long outbursts of these sources will generate enough heat in the crust that it will become strongly out of equilibrium with the core and it might take years to even decades (depending on crust properties) for the crust to cool down and reach equilibrium with the core again.

\begin{table*}[htb]
\fontsize{8}{10}\selectfont
\caption{The sample of crust cooling sources}\label{table:cooling}
\begin{tabular}{lccl}
\topline
Sources                & Instrument$^a$ & Remark   & References\\\midline
\multicolumn{2}{l}{\underline{Quasi-persistent sources}}\\
MAXI J0556-332$^b$     & C, S, X        &          & \citet{2014ApJ...795..131H}\\
EXO 0748-676           &  C, S, X          & Eclipser & \citet{2009MNRAS.396L..26D,2011MNRAS.412.1409D,2014ApJ...791...47D,2011AA...528A.150D}\\
                       &                &          & \citet{2017arXiv170604784C}\\
MXB 1659-298           & C, X           & Eclipser & \citet{2003ApJ...594..952W,2004ApJ...606L..61W,2006MNRAS.372..479C,2008ApJ...687L..87C,2013ApJ...774..131C}\\
XTE J1701-462$^b$      &  C, X          &          & \citet{2010ApJ...714..270F,2011ApJ...736..162F}\\
KS 1731-260            & C, X           &          & \citet{2001ApJ...560L.159W,2002ApJ...573L..45W,2002ApJ...580..413R}\\
                       &                &          & \citet{2006MNRAS.372..479C,2010ApJ...722L.137C,2016ApJ...833..186M}\\
HETE J1900.1-2455$^c$  & C              &          & \citet{2017MNRAS.465L..10D}\\
\hline
\multicolumn{2}{l}{\underline{Ordinary transients}}\\
IGR J17480-2446        & C               & In Terzan 5          & \citet{2011MNRAS.414L..50D, 2011MNRAS.418L.152D,2013ApJ...775...48D,2015MNRAS.451.2071D}\\
Swift J174805.3-244637 & C               & In Terzan 5          & \citet{2015MNRAS.451.2071D}\\
1RXS J180408.9-342058$^c$  & S, X              &            & \citet{2017MNRAS.tmp...16P}\\
Aql X-1$^{b,c}$ (4U 1908+00)    & S              &          & \citet{2016MNRAS.456.4001W}\\
\hline
\end{tabular}
\tablenotes{$^a$ C: {\it Chandra}; S: {\it Swift}; X: {\it XMM-Newton}}
\vspace{-0.4cm}
\tablenotes{$^b$ Small accretion flares have been observed during the crust cooling phase.}
\vspace{-0.4cm}
\tablenotes{$^c$ Crust cooling still needs to be confirmed as mechanism behind observed behaviour.}
\end{table*}

\subsection{Quasi-persistent sources} 

In 2001, KS 1731-260 turned off after having accreted for $\sim$12.5 years. Soon after the end of this transition, \citet[][]{2001ApJ...560L.159W} obtained a {\it Chandra} observation to study the quiescent properties of the source. The results of this observation, together with the first crust cooling models presented by \citet[][]{2002ApJ...580..413R}, marked the start of the research field that uses the cooling of accretion-heated NS crusts to investigate the properties of such crusts. Since then, the source has indeed been found to display crust cooling and this cooling has now been followed for $\sim$15 years \citep[see Figure \ref{fig:coolingcurves}, left;][]{2002ApJ...573L..45W,2006MNRAS.372..479C,2010ApJ...722L.137C,2016ApJ...833..186M}. Also in 2001, MXB 1659-298 turned off after having accreted for $\sim$2.5 years. For this source crust cooling was also observed after its outburst and it has now been followed for over a decade as well \citep[Figure \ref{fig:coolingcurves}, left;][]{2003ApJ...594..952W,2004ApJ...606L..61W,2006MNRAS.372..479C,2008ApJ...687L..87C,2013ApJ...774..131C}. 

Several theoretical studies have tried to model these two sources and we refer to these papers for the details of what can be inferred from the cooling curves \citep[][]{2002ApJ...580..413R,2007MNRAS.382L..43S,2009ApJ...698.1020B,2012arXiv1201.5602P,2013PhRvL.111x1102P,2015A&A...577A...5T,2016MNRAS.461.4400O}. One of the main conclusions was that since both sources cooled down rapidly in quiescence, they must have a high heat conductivity in their NS crusts.  Originally, this was a surprising result because thermonuclear burning of the accreted matter produces a wide range of different elements. This was expected to cause a rather impure and disordered structure in the crust, resulting in a low thermal conductivity of the crust \citep[e.g., see][]{2001NuPhA.688..150S}. Later on, however, it was found that chemical separation is likely to occur
during crystallisation, resulting in a high crustal conductivity \citep[][]{2007PhRvE..75f6101H}. Hence, observations of these two sources and their interpretation have 
provided us with the first observational evidence that the crust of an accreting NS
forms a crystal and not just an amorphous solid.

The very long cooling curves for both sources (that are now available) allow to probe the properties of the deepest layers in the crust \citep[e.g., the layers where the ions become highly non-spherical due to the very high densities or even as deep as the so-called pasta layer;][]{2015PhRvL.114c1102H,2017ApJ...839...95D}. In addition, for both sources the observed surface temperatures are very low, indicating that their cores are very cold, which puts tight constraints on their heat capacity and therefore the degree of superfluidity of the core matter and its composition \citep[][]{2017PhRvC..95b5806C}.

Additional cooling observations of both sources could in principle further constrain the core properties, but for KS 1731-260 the most recent data show that the crust and core are back in thermal equilibrium again and therefore we are now probing the core properties already \citep[][]{2016ApJ...833..186M}. Further cooling observations  to probe the core physics of MXB 1629-928 are not possible anymore since this source went back into outburst in 2015 \citep[][]{2015ATel.7943....1N}. However, very recently (March 2017), this source turned off again \citep[][]{2017ATel10169....1P} and several quiescent observations will be performed in the time period 2017-2019 to study the crust cooling of this source after a second outburst. Since this last outburst lasted only $\sim$1.5 years (compared to the $\sim$2.5 year of the previous outburst), it is plausible that differences in the crust cooling behaviour can be observed compared to what was seen after the previous outburst. This will allow the possibility to separate the effects caused by differences in outburst behaviour (i.e., duration, variability) from intrinsic NS properties (i.e., mass, radius, thermal properties of the crust). 

Besides KS 1731-260 and MXB 1659-298, for three other quasi-persistent NS transients crust cooling curves have been obtained: XTE J1701-462, EXO 0748-676, and MAXI J0556-332 (Figure \ref{fig:coolingcurves}; Table \ref{table:cooling}). The quiescent monitoring campaign on XTE J1701-462 (a very bright source that was active for $\sim$1.6 years until August 2007) has shown that the source behaviour is complex \citep[Figure \ref{fig:coolingcurves}, left;][]{2010ApJ...714..270F,2011ApJ...736..162F}. Apart from several minor accretion flares, the source first exhibited rapid cooling, suggesting that this source also must have a high crustal conductivity. Later on, the source reached a plateau which suggested that the crust had reached thermal equilibrium with the core but the most recent observations showed that the temperature significantly decreased further (see Figure \ref{fig:coolingcurves}, left). Such a drop is consistent with one of the curves that was calculated for this source \citep[][]{2012arXiv1201.5602P,2013PhRvL.111x1102P}; i.e., the curve for which a cold core was assumed which would require enhanced neutrino emission from the core (the cooling curve of this source was also modelled by \citet[][]{2014ApJ...783L...3M} and \citet[][]{2015A&A...577A...5T}). 

The monitoring campaign on EXO 0748-676 (a relatively faint source but with an outburst duration of $\sim$24 years that ended in 2008) has shown that also for this source the NS crust initially cooled down relatively fast, but very quickly after that it slowed down and the most recent data show that the source is still only slowly cooling \citep[Figure \ref{fig:coolingcurves}, left;][]{2009MNRAS.396L..26D,2011MNRAS.412.1409D,2014ApJ...791...47D}. So far the source has not yet cooled as dramatically as the other sources which is remarkable since it was in outburst for $\sim$24 years. This suggest that its core temperature is very high and also it is likely that the source has a very high accretion duty cycle.

In 2012, MAXI J0556-332 turned off after exhibiting a very bright outburst of $\sim$1.5 years. These outburst properties were very similar as those observed for XTE J1701-462 from which one would infer that a similar amount of energy should have been generated in the NS crust in both sources. However, the NS crust in MAXI J0556-332 was very hot compared to what has been observed for XTE J1701-462 \citep[Figure \ref{fig:coolingcurves}, right;][]{2014ApJ...795..131H}. Although unexpected, such a very hot crust offered a chance to test the importance of a new, very efficient neutrino cooling mechanism in the crust discovered by \citet[][]{2014Natur.505...62S}. So far no evidence could be found that this mechanism was indeed active in MAXI J0556-332 \citep[][]{2015ApJ...809L..31D,2017ApJ...837...73M}.

To obtain the very high temperatures in the crust observed in MAXI J0556-332, a lot of energy has to have been generated at shallow depths in the crust ($\sim$150 meters) due to an unknown heating source; much larger than what can be accounted for by the standard nuclear reaction processes (the ``shallow-heating'' problem). Although a similar shallow-heating source was inferred for other cooling sources and from other studies  involving accreting NSs \citep[i.e., studying thermonuclear bursts; see, e.g.,][]{2006ApJ...646..429C,2012A&A...547A..47I,2012ApJ...748...82L}, the amount of heating needed in MAXI J0556-332 is an order of magnitude larger than previously required. The origin of this shallow-heat source is not understood and we refer to \citet[][]{2015ApJ...809L..31D} for a discussion.

About 200 days in the cooling phase, the source exhibited a small additional outburst \citep[][]{2014ApJ...795..131H}, but this outburst did not appear to have a measurable effect on the cooling curve (see Figure \ref{fig:coolingcurves}, right) and the shallow-heating mechanism appeared not to be active during this outburst \citep[][]{2015ApJ...809L..31D}. Unexpectedly, the source exhibited another outburst in early 2016 \citep[][]{2016ATel.8513....1N}. Since this new outburst was bright and lasted several months, an excellent opportunity arose to investigate the expected reheating of the crust caused by the accretion of matter during this new outburst. Surprisingly, it was found that the crust was indeed reheated by this outburst \citep[Figure \ref{fig:coolingcurves}, right;][]{parikh0556} but only by a very small amount, indicating that the shallow-heating mechanism was active during this outburst but only at a much reduced level (although it had to be active to explain the amount reheating caused by the third outburst which cannot be explained by deep crustal heating only). The reason for this different behaviour is unclear but it demonstrates that obtaining a second cooling curve of the same source provides very valuable information and allows to separate the effects of variability in outburst behaviour from intrinsic NS properties.

Apart from the crust cooling curves observed for the above five discussed quasi-persistent sources, a quiescent observation was also obtained for HETE J1900.1-2455 $\sim$ 180 days after the end of its $\sim$10 years outburst \citep[][]{2017MNRAS.465L..10D}. Unfortunately, earlier crust cooling observations were not obtained and so far only one quiescent observation is available. Therefore, it is not possible to determine if the source indeed exhibits crust cooling similar to the other systems. The available observation showed a cold NS, significantly cooler than the other sources when they were also half a year in quiescence. However, the source was not very bright in outburst and this cold crust could be explained using typical parameters observed for the other targets \citep[][]{2017MNRAS.465L..10D}. An additional {\it Chandra} observation of this source is scheduled for 2018 with which we will be able to determine if the source indeed exhibited a heated crust after the end of its outburst, or that the crust and core were already in equilibrium only $\sim$180 days after the outburst was over.

\subsection{Ordinary transients}\label{section:ordinary}

It was already predicted by \citet[][]{1998ApJ...504L..95B} that for certain subgroups of ordinary transients the cooling of the crust might be observable as well after the end of one of their outbursts. The best candidates would be those systems that have a relatively cold core (and thus a relatively low base luminosity) and exhibit bright outbursts (with luminosities exceeding $10^{37}$ erg s$^{-1}$). Only for those systems the relatively minor increase in crustal temperatures (due to the shorter outbursts) might be detectable above the base level and also only for a short period of time (at most weeks to months) before the crust would be in equilibrium with the core again. However, when this hypothesis was tested by monitoring observations of ordinary transients after their outbursts, it turned out that crust heating and cooling could easily be detected as well. Several sources have now been studied in detail.

IGR J17480-2446 and Swift J174805.3-244637 are two ordinary transients located in the globular cluster Terzan 5 which both exhibited an approximately two month bright (peak X-ray luminosity of $5\times 10^{37} - 10^{38}$ erg s$^{-1}$) outbursts in 2010 and 2012, respectively \citep[][]{2010ATel.2952....1A,2014ApJ...780..127B}. Since {\it Chandra} had frequently observed this cluster when these sources were not in outburst, their base quiescent levels were known before they exhibited their first known outbursts. IGR J17480-2446 has a relatively cold base level \citep[][]{2011MNRAS.412L..68D} but Swift J174805.3-244637 is significantly hotter \citep[][]{2014ApJ...780..127B}. However, for both sources it could be determined that even a two-month outburst was enough to heat the crust out of equilibrium with the core so that we could observe the crust cooling \citep[][]{2011MNRAS.414L..50D,2013ApJ...775...48D,2015MNRAS.451.2071D}. Due to the higher base level of Swift J174805.3-244637 the crust was again in equilibrium with the core within $\sim$200 days, but the crust cooling of IGR J17480-2446 is ongoing \cite[Figure \ref{fig:coolingcurves}, left;][]{2015MNRAS.451.2071D}. To explain the significant heating of the NS crust in IGR J17480-2446, about 1-2 MeV per nucleon additional heat should have been generated in its crust due to the shallow-heat mechanism. No such additional heat source in the crust was required for the NS in Swift J174805.3-244637 \citep[although it could not be excluded either; see][]{2015MNRAS.451.2071D}.

Apart from the two transients in Terzan 5, for two additional ordinary transients potential crust cooling has been observed. It was found that the {\it Swift} data obtained in quiescence for the frequently recurring transient Aql X-1  could be explained by a cooling crust \citep[][]{2016MNRAS.456.4001W}. However, this needs to be confirmed to determine if indeed crust cooling is observed in this source and not some other process that could also cause a gradual decay of the observed luminosities (i.e., low level accretion). Assuming that crust cooling is observed in this system, then this source would be an excellent target to investigate the effects of different outburst properties on the crust heating and cooling behaviour \citep[][]{ootes2017}. In addition, very recently strong evidence for crust cooling was observed in 1RXS J180408.9-342058, which turned off after an accretion episode of $\sim$4.5 months in 2015 \citep[][]{2017MNRAS.tmp...16P}. But also for this source this needs to be confirmed and additional {\it XMM-Newton} observations are scheduled to determine its further evolution in quiescence. If indeed crust cooling was observed in Aql X-1 and 1RXS J180408.9-342058, then the shallow-heating mechanism should have been present in both sources to heat their crusts to the observed temperatures \citep[][]{2016MNRAS.456.4001W,2017MNRAS.tmp...16P,ootes2017}.
 
\section{High-magnetic field NS systems} \label{section:magnetic}

Until recently, nearly all the efforts to study crust and core cooling of accretion-heated NSs had focussed on those NSs that have very weak magnetic fields  ($B \sim10^{8-10}$ G). In such systems, the fields are so weak that they have very little effect on the heating and cooling of the NS crusts,  making modelling of the cooling data significantly less complex. However, this excludes a large fraction of the known accreting NSs. Therefore, recently several investigations have been performed to study the strongly magnetized NSs (B $\sim 10^{12-13}$ G) as well to determine the effect of the magnetic field on the heating and cooling mechanisms.  

These studies have focused so far on the Be/X-ray transients in which a magnetized NS is orbiting a Be star in an eccentric orbit (although the NSs might also be visible in quiescence in the supergiant fast X-ray transients; see, e.g., \citealt[][]{2005A&A...441L...1I}). Periodically the NSs come close to their Be companions and if decretion disks are present and these are large enough, the NSs can penetrate these disks causing the accretion of matter during so-called type-I outbursts. Such outbursts are relatively faint, with peak luminosities of $10^{36-37}$ erg s$^{-1}$. Although the NS crusts and cores should be heated during these outbursts as well, the crust cooling studies (see below) in Be/X-ray transients have focused on the type-II outbursts which are much brighter (reaching luminosities of $10^{38-39}$ erg s$^{-1}$; the mechanism behind such bright outbursts is not understood; see \citet[][]{2011Ap&SS.332....1R} for a review of Be/X-ray transients).

\begin{table*}[htb]
\fontsize{8}{10}\selectfont
\caption{Reports on observations of quiescent Be/X-ray transients$^a$}\label{table:BeX}
\begin{tabular}{lccl}
\topline
Sources                & Instrument$^b$ & Remark      & References\\\midline
IGR J01363+6610        & C, X           &             & \citet[][]{2008ApJ...685.1143T,2011ApJ...728...86T}\\
4U 0115+63             & B, S, X        &  Pulsations & \citet[][]{2001ApJ...561..924C,2002ApJ...580..389C,2016MNRAS.463L..46W}\\
                       &                &  Cooling?   & \citet[][]{2017arXiv170400284R,2017arXiv170304634T}\\
V 0332+53              & B, C, X        &  Cooling?   & \citet[][]{2002ApJ...580..389C,2016MNRAS.463L..46W}\\
                       &                &             & \citet[][]{2016MNRAS.463...78E,2017arXiv170304634T}\\
1A 0535+26             & B, EX,         &  Pulsations & \citet[][]{2000AA...356.1003N,2004NuPhS.132..476O,2005AA...431..667M}\\
                       & RX, X          &             & \citet[][]{2013ApJ...770...19R,2014AA...561A..96D}\\
MXB 0656-072           & C              &             & \citet[][]{2017arXiv170304634T}\\
4U 0728-25             & S              &             & \citet[][]{2017arXiv170304634T}\\
RX J0812.4-3114        & C              &             & \citet[][]{2017arXiv170304634T}\\
GS 0834-43             & C              &             & \citet[][]{2017arXiv170304634T}\\
GRO J1008-57$^c$       & C, S           &             & \citet[][]{2017arXiv170304528T,2017arXiv170304634T}\\
4U 1118-615            & C              &  Pulsations & \citet[][]{2007ApJ...658..514R}\\
4U 1145-619            & E              &  Pulsations & \citet[][]{1987ApJ...312..755M,2007ApJ...658..514R}\\
2S 1417-624            & C              &             & \citet[][]{2017arXiv170304634T}\\
2S 1553-542            & C              &             & \citet[][]{2017arXiv170304634T}\\
Swift J1626.6-5156     & C              &  Pulsations & \citet[][]{2011MNRAS.415.1523I,2017arXiv170304634T}\\
XTE J1829-098          & C, X           &             & \citet[][]{2007ApJ...669..579H}\\
GS 1843+00             & C              &             & \citet[][]{2017arXiv170304634T}\\
XTE J1946+274          & C              &  Pulsations & \citet[][]{2015AA...582A..53O,2017arXiv170304634T}\\
KS 1947+300            & C              &  Pulsations & \citet[][]{2017arXiv170304634T}\\
EXO 2030+375           & S              &             & \citet[][]{2017arXiv170706496F}\\
GRO J2058+42          & C              &             & \citet[][]{2005ApJ...622.1024W}\\
SAX J2103.5+4545       & C              &  Pulsations & \citet[][]{2014MNRAS.445.1314R,2017arXiv170304634T}\\
Cep X-4 (GS 2138+56)   & C, R           &  Pulsations & \citet[][]{1995AA...295..413S,2017arXiv170304634T}\\
SAX J2239.3+6116       & C              &             & \citet[][]{2017arXiv170304634T}\\
\hline
\end{tabular}
\tablenotes{$^a$ A similar table was also presented by \citet[][]{2014MNRAS.445.1314R}.}
\vspace{-0.4cm}
\tablenotes{$^b$ B: {\it BeppoSAX}; C: {\it Chandra}; E: {\it Einstein}; EX: {\it EXOSAT}; R: {\it ROSAT}; RX: {\it RXTE}; S: {\it Swift}; X: {\it XMM-Newton} }
\vspace{-0.4cm}
\tablenotes{$^c$ Unclear if GRO J1008-47 was observed in true quiescence since it was relatively bright ($10^{34-35}$ erg s$^{-1}$).}
\end{table*}

A growing number of Be/X-ray transients has been observed in quiescence long after their last outbursts (Table \ref{table:BeX}). Strong evidence exists that for some systems very low-level accretion of matter continues onto the NS magnetic poles \citep[e.g.,][]{2013ApJ...770...19R,2014AA...561A..96D,2017arXiv170304634T}. For other systems, the X-rays could indeed be due to cooling emission from an accretion-heated NS \citep[e.g.,][]{2001ApJ...561..924C,2002ApJ...580..389C,2014MNRAS.445.1314R,2016MNRAS.463L..46W,2016MNRAS.463...78E,2017arXiv170304634T}. The latter systems are prime targets to investigate the effect of a strong magnetic field on the heating and cooling processes. So far these NSs do not appear to behave differently from the weakly magnetized NS systems \citep[e.g.,][]{2016MNRAS.463...78E,2017arXiv170304634T}. 

Important for the interpretation of the cooling emission in Be/X-ray transients is the fact that it is possible that the observed cooling emission might arise from hot spots at the NS magnetic poles \citep[see][]{2016MNRAS.463L..46W,2017arXiv170400284R}. However, the rest of the surface could contribute (or dominate) significantly to the cooling radiation, but its lower temperature might make this contribution unobservable \citep[see discussion in][]{2016MNRAS.463...78E}. In addition, owing to the young age of Be/X-ray transients, it is possible that the NS crusts in these systems have not been fully replaced by accreted matter and a hybrid crusts might be present (partly accreted, partly original matter). Not all accretion-induced nuclear reactions might occur in such a crust, causing less heat generation and making the NS potentially very cold \citep[][]{2013MNRAS.432.2366W}.

Similar to the transient NS LMXBs, if the outbursts in Be/X-ray transients are very bright or last long (i.e., the type-II outbursts), the NS crusts in these systems might be lifted out of thermal equilibrium with the cores during the outbursts. Very recently \citet[][]{2016MNRAS.463L..46W} performed such studies for 4U 0115+63 and V 0332+53 and found strong evidence for crust heating and cooling in these two systems. However,  \citet[][]{2017arXiv170400284R} presented follow-up observations for 4U 0115+63 and showed that the situation is complex and additional assumptions (i.e., about the configuration of the magnetic field in the crust) have to be made to be able to explain the observations using the crust cooling hypothesis. The same study found that the crust cooling phase in 4U 0115+63 lasted at most $\sim$250 days, which is  much shorter than typically seen for the weak field NS systems. Similar short ($<$1 year) crust cooling phases were indicated for some sources (i.e., GS 0834-430) presented in the paper of \citet[][]{2017arXiv170304634T}. More studies have to be performed to confirm the crust cooling phase in Be/X-ray transients and to determine why it lasts for such a short time in these systems.

\section{Conclusion}

Clearly, over the last 15 years the cooling studies of accretion-heated NSs have made considerable progress, leading to important new insights in the properties of NSs and how they react to the accretion of matter. However, many unresolved issues remain, both observationally as well as in the theoretical models. One particular urgent problem is the mechanism(s) behind the shallow heating that is needed in most crust cooling sources to explain their behaviour. Observing multiple outbursts for several sources will be of significant help in determining the nature of this shallow-heat mechanism. However, the variations in the accretion rate during these outbursts have to be properly taken into account as well as shown by \citet[][]{2016MNRAS.461.4400O,ootes2017}

For the core cooling studies, it is important to create a source sample in which the data analysis has been performed homogeneously and using coherent estimates of $\langle \dot{M} \rangle$. Currently, we are in the process of updating Figure \ref{fig:lxversusmdot}, taking into account all known uncertainties and adding all available quiescent systems (see Tables \ref{table:qsources}-\ref{table:cooling} for the list of sources which is significantly larger than the number of sources shown in Figure \ref{fig:lxversusmdot}).  Also other effects have to be taken into account such as the effects of variations in the chemical composition of the envelope \citep[][]{2017arXiv170208452H} and the possibility of having a large fraction of the observed emission originating from possible hotspots instead of the whole surface \citep[][]{2016ApJ...826..162E}.

High magnetic field NS transients will be very useful to investigate the effects of the strong magnetic field on the heating and cooling of the NS crust and core. However, it will likely remain difficult to determine the exact amount of cooling emission if hotspots are present \citep[][]{2016MNRAS.463...78E} and to disentangle magnetic field effects from those induced by a potential hybrid NS crust \citep[][]{2013MNRAS.432.2366W}. Crust cooling studies using these systems will be very valuable, but it is clear that the situation is more complex than what has been observed for the low-field NS systems \citep[][]{2017arXiv170400284R}. In this context, it might be prudent to compare the accreting systems with magnetars, for which crust cooling might also have been observed \citep[e.g.,][]{2002ApJ...580L..69L,2012ApJ...750L...6P,2012ApJ...754...27R,2012ApJ...761...66S,2014ApJ...786...62S,2013ApJ...770...65R,2015MNRAS.449.2685C}.

\section*{Acknowledgement}

RW acknowledges support from a NWO Top Grant, Module 1. ND is supported by a NWO VIDI grant. DP’s work is supported by grant number 240512 from Conacyt CB-2014-1. We thank Aastha Parikh for making the most up to date version of Figure \ref{fig:coolingcurves}. We thank Laura Ootes, Aastha Parikh,  Alicia Rouco Escorial, and Liliana Rivera Sandoval for comments on an earlier version of this review.







\end{document}